# Ballistic Dynamics of Flexural Thermal Movements in a Nano-membrane Revealed with Subatomic Resolution

Tongjun Liu[1], Jun-Yu Ou[1], Nikitas Papasimakis[1], Kevin F. MacDonald[1], Vitalyi E. Gusev[2], and Nikolay I. Zheludev[1, 3]

[1] *Optoelectronics Research Centre and Centre for Photonic Metamaterials, University of Southampton, Highfield, Southampton, SO17 1BJ, UK*

[2] *Laboratoire d'Acoustique de l'Université du Mans (LAUM), Institut d'Acoustique-Graduate School (IA-GS), CNRS, Le Mans Université, 72085 Le Mans, France*

[3] *Centre for Disruptive Photonic Technologies, SPMS and TPI, Nanyang Technological University, Singapore, 637378, Singapore*

Flexural oscillations of free-standing films, nano-membranes and nano-wires are attracting growing attention for their importance to the thermal, electrical and mechanical properties of 2D materials. Here we report on the observation of short-timescale ballistic motion in the flexural mode of a nano-membrane cantilever, driven by thermal fluctuation of flexural phonons, including measurements of ballistic velocities and displacements performed with sub-atomic resolution, using a new free electron edge-scattering technique. Within intervals <10 µs, the membrane moves ballistically at a constant velocity, typically ~300 µm/s, while Brownian-like dynamics emerge for longer observation periods. Access to the ballistic regime provides verification of the equipartition theorem and Maxwell-Boltzmann statistics for flexural modes, and can be used in fast thermometry and mass sensing during atomic absorption/desorption processes on the membrane. We argue that the ballistic regime should be accounted for in understanding the electrical, optical, thermal and mechanical properties of 2D materials.

Flexural deformations and modes of oscillation are now understood to be of fundamental importance to the thermal, optical, electrical and mechanical properties of graphene, other 2D materials [1-5], and to the optical properties of photonic metamaterials through near-field coupling among resonators and mechanochromic effects [6, 7]. In contrast with a classical Brownian particle in a fluid that is thermally perturbed by *external* collisions with ambient atoms, under vacuum thermal movements are driven *internally* by momentum transfer from the annihilation and creation of the flexural phonons.

As long ago as 1906, Einstein realized that the commonly held picture of diffusive thermal motion, characterized by erratic, discontinuous changes in speed and direction, must break down at short time and length scales, where inertia becomes significant [8] – objects must move ballistically between 'collision' events. He concluded that this regime of motion would be impossible to observe as to do so would require, at the time, unimaginably high spatial and temporal measurement resolution. Indeed, even today, phonon-dominated dynamics in free-standing films, nano-membranes, nano-wires and cantilevers remained underexplored because there were no technologies available for quantifying their short-timescale nano/picoscale motion.

We show here that detection of variations in secondary electron emission from the edge of a moving (oscillating) nano-membrane interrogated with a focused electron beam provides for measurements of the membrane's position with nanosecond temporal resolution and sub-atomic displacement sensitivity. The detection method for the first time reveal the Einstein-predicted ballistic regime of thermal flexural motion of the membrane in short time scales. Our experiments allow the measurement of velocities of consecutive steps of membrane movement and their statistics and provide the first direct experimental verification of the applicability of the equipartition theorem and Maxwell-Boltzmann statistics to flexural dynamics.

We investigated the dynamics of thermal motion in the out-of-plane flexural mode of a cantilever cut from a free-standing gold nano-membrane. The choice of the material and cantilever geometry being dictated by considerations of its effective mass, and the natural frequency and quality factor of its fundamental flexural mode, and secondary electron yield, to facilitate observation of the ballistic regime. It was 30 nm thick and 62 µm long with a width tapered from 0.6 µm at the fixed end to 3 µm at the other, and an effective mass $m_{eff}$ = 47 pg (see Supplementary Information). The secondary electron flux generated by scattering of an electron beam tightly focused on the edge of the membrane is highly sensitive to its position (see Materials and Methods). Position measurements as a function of time reveal Gaussian distributions of the membrane's position and velocity with root-mean-square values of $\langle x \rangle$= 8.3 nm and $\langle v \rangle$= 0.30 mm/s, respectively (Fig. 1). The natural oscillation frequency of the cantilever $\omega_0/2\pi$ = 5.7 kHz and the damping time $\tau_b = 1/\gamma$ = 14 ms were evaluated from the Fourier spectrum of the displacement autocorrelation function.

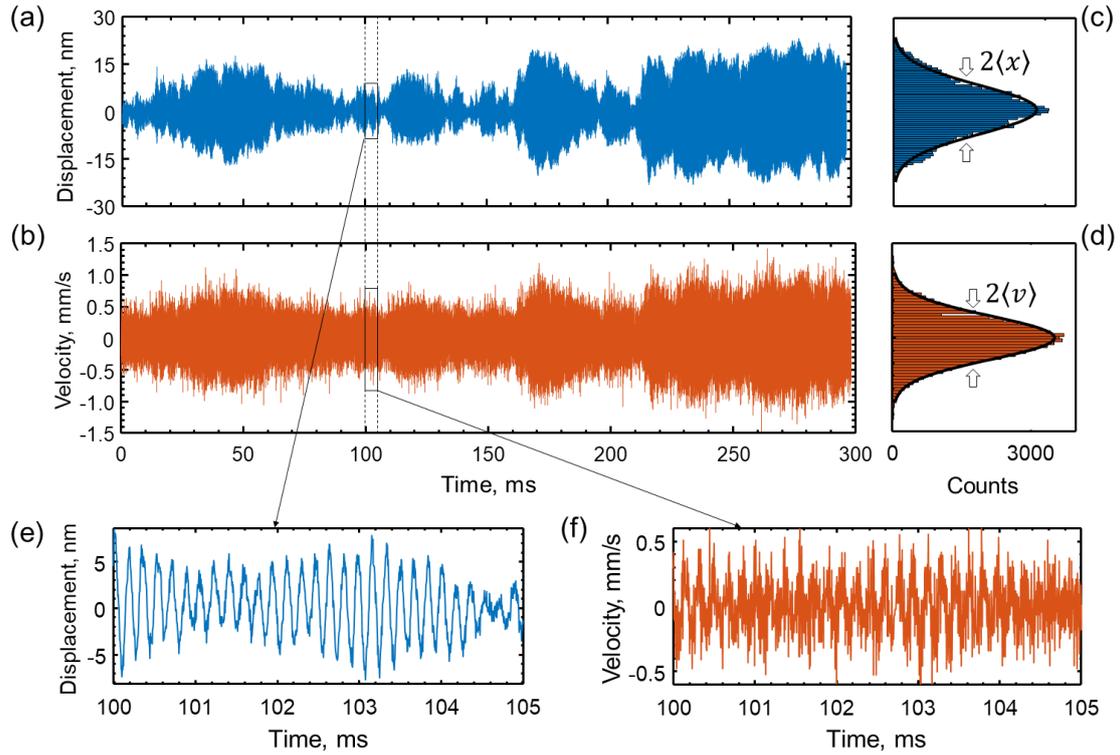

**Fig. 1. Thermomechanical motion of a gold nano-membrane cantilever measured by free electron edge scattering.** Time series recording of displacement (a) and corresponding [derived] velocity (b) of the tip of a cantilever moving in its fundamental flexural mode under vacuum. (c) and (d) show corresponding displacement and velocity distributions. Overlaid black lines are Gaussian fittings. (e) and (f) show zoomed in sections of (a) and (b) respectively, in which the oscillatory period of the mode $\theta = 1.75 \times 10^{-4}$ s is resolved.

To reveal the detailed nature of nano-membrane cantilever thermal motion, from the experimental data we evaluate mean squared displacement $\langle \delta x(\tau)^2 \rangle$ as a function of observation time $\tau$ (Fig. 2a). For small observation intervals $\tau \ll \theta$ (where $\theta = 1.75 \times 10^{-4}$ s is the oscillation period of the cantilever) the mean squared displacement $\langle \delta x(\tau)^2 \rangle$ grows quadratically with $\tau$. This is direct evidence of the ballistic motion regime, as it means that velocity is constant over the observation interval. This is confirmed by the normalized velocity autocorrelation function $\langle v(t)v(t+\tau) \rangle / (k_B T/m_{eff})$ plotted in Fig. 2b, which evaluates how close the velocity at the end of the observation period is to the velocity at the beginning. At short time intervals, the near-unity value of the autocorrelation function is again explicit evidence of the ballistic regime. From Fig. 2a one can conclude that over intervals up to ~$10^{-5}$s the cantilever moves ballistically over average distances up to 3 nm.

In essence, the ballistic regime implies that within a short observation interval $\tau \ll \theta$, the natural oscillation of the cantilever is not significantly disturbed by momentum transfer related to the annihilation and creation of individual flexural phonons. At room temperature ($T$ = 300 K) the average number of thermal phonons [9, 10] with energy $\hbar\omega$ in the flexural mode is $\bar{n}_{th} \approx k_B T/\hbar\omega = 1.1 \times 10^9$, while the average lifetime of the flexural phonons can be evaluated as $(\bar{n}_{th}\gamma)^{-1}$ = 13 ps, during which time the cantilever moves an average distance of $\langle v \rangle / (\bar{n}_{th}\gamma)$ = 3.9 fm. We observe that phonon momentum transfer events begin to affect the ballistic regime of natural oscillation only when the observation intervals $\tau > 10^{-5}$ s, i.e., about 6% of the oscillation period $\theta$. During this period of observation $\bar{n}_{th}\gamma\tau$ ~$10^6$ phonons out of the ~$1.1 \times 10^9$ in the mode are created and dissipated.

For longer observation periods, $\langle \delta x(\tau)^2 \rangle$ becomes a complex oscillating function of $\tau$. A truly diffusive regime, wherein $\langle \delta x(\tau)^2 \rangle \propto \tau$ and which is characteristic of free particle movement, is not observable in cantilevers, where the dynamics are affected by a restoring force. The statistical properties of cantilever thermal motion are described by the Langevin model [11, 12] for a harmonic oscillator: $\ddot{x} + \gamma\dot{x} + \omega_0^2 x = F_T(t)/m_{eff}$, where $F_T(t) = \sqrt{2k_B T\gamma/m_{eff}}\eta(t)$ is the thermal force related to the dissipation factor $\gamma$ through the fluctuation-dissipation theorem [13], $k_B$ is the Boltzmann constant, and $\eta(t)$ is a delta-correlated normalized white noise term: $\langle \eta(t) \rangle = 0$; $\langle \eta(t)\eta(t') \rangle = \delta(t-t')$. There is a remarkably good correlation between experimental data and the Langevin model over the entire range of observation times, particularly at $\tau \ll \vartheta$ where the value of $\langle \delta x(\tau)^2 \rangle$ accurately follows the $\tau^2 k_B T/m_{eff}$ dependence derived from the model.

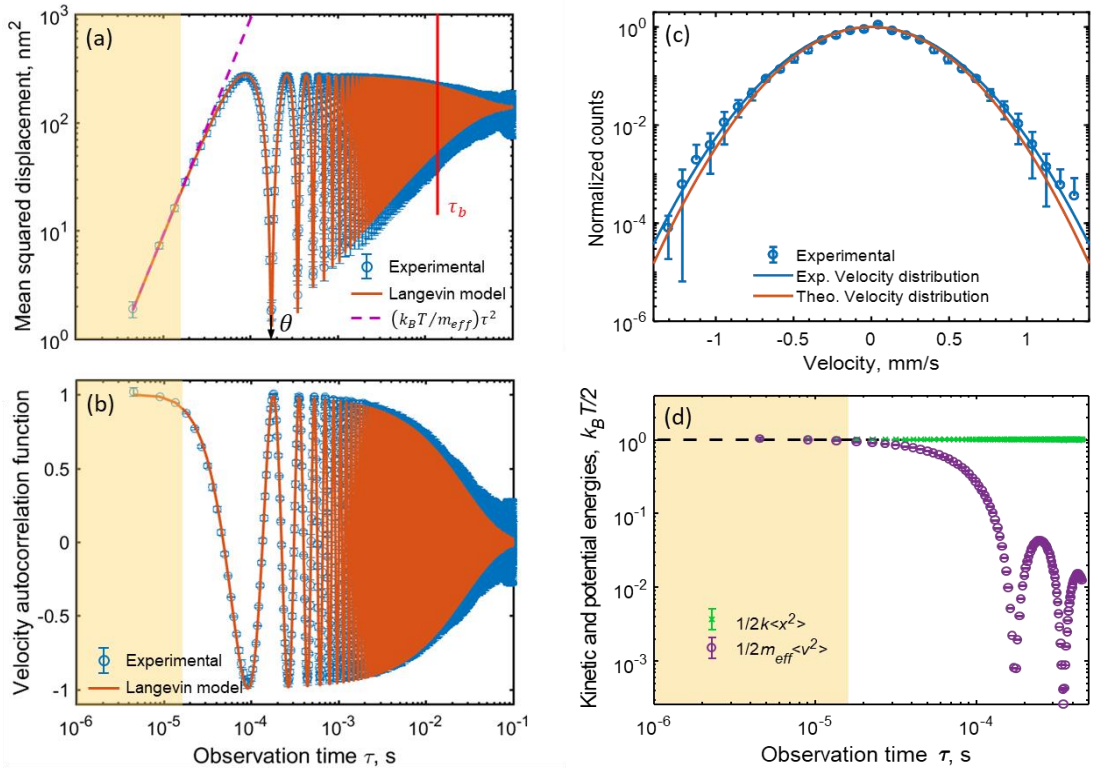

**Fig. 2. Statistics of nano-membrane cantilever thermal motion.** (a) Mean squared displacement $\langle \delta x(\tau)^2 \rangle$ of the membrane cantilever tip as a function of the observation time interval $\tau$. Experimental data are plotted as blue circles; The orange line is derived from the Langevin model for thermal motion of a harmonic oscillator; The violet dashed line is an analytical asymptote for ballistic motion at a constant velocity of $\sqrt{k_B T/m_{eff}}$ = 0.297 mm/s. (b) Normalized velocity autocorrelation function $\langle v(t)v(t+\tau) \rangle /(k_B T/m_{eff})$ as a function of the observation time interval. (c) Measured cantilever tip velocity distribution for an observation time interval $\tau$ = 4.7 μs [blue circles]. Solid lines are Maxwell-Boltzmann distributions: blue, as a best-fit to the experimental data, with $v_{rms}$ = 0.303 mm/s; Orange, with $v_{rms}$ = 0.297 mm/s from equipartition theorem. Experimentally measured values of $\frac{1}{2}k\langle x^2 \rangle$ [green symbols and line] and $\frac{1}{2}m_{eff}\langle v^2 \rangle$ [purple] as functions of the observation time $\tau$. The former equates to the potential energy of the cantilever. At short timescales, where $v$ represents the well-defined ballistic velocity, the latter equates to kinetic energy. Yellow shaded zones in (a), (b) and (d) denote the ballistic regime.

In the ballistic regime of motion ($\tau \ll \theta$) nanomembrane cantilever velocity $v$ is well defined, so the distribution of velocities over an ensemble of sampling events can be established, as plotted in Fig. 2c for the shortest observation interval $\tau$ = 4.7 μs. A Maxwell-Boltzmann distribution fitting to the data yields a root-mean-square velocity $\langle v \rangle$ = 0.303 mm/s that is well-matched to the value obtained from energy equipartition theorem: $\sqrt{k_B T/m_{eff}}$ = 0.297 mm/s. The small 2% discrepancy between these values is related to the shot noise of the electron beam current in the experimental case. Compliance with the equipartition theorem, which stipulates that the Boltzmann energy $k_B T$ should be equally distributed between potential and kinetic energies of the cantilever is also illustrated in Fig. 2d. Here we plot experimental values of $\frac{1}{2}k\langle x^2 \rangle$ and $\frac{1}{2}k\langle v^2 \rangle$ as functions of the observation time interval $\tau$. The former equates to the potential energy of the cantilever; the latter equates to kinetic energy *only* at short intervals $\tau \ll \theta$, where $v$ represents the well-defined ballistic velocity. Convergence of the two traces at short timescales thereby confirms equipartition of kinetic and potential energies in the ballistic regime (the yellow shaded band in Fig. 2d).

Our observations of the thermomechanical motion of a nano-membrane cantilever reveal the following dynamics: At short timescales, up to about 10 μs the membrane moves with constant velocity (i.e. ballistically), with the average displacement being directly proportional to the observation time interval. Brownian-like dynamics emerge for longer observation times, when membrane motion is caused by multiple phononic creation/annihilation events: average displacement grows as the square root of time. (The emission and absorption of thermal photons makes a negligible contribution due to their low momentum.)

When length of the observation interval becomes equal to the natural oscillation period, average displacement reaches a minimum. For intervals much longer than the oscillation period, the mean squared displacement $2k_BT/(m_{eff}\omega_0^2)$ is proportional to temperature and is independent of observation time. High sampling-rate measurements of the instantaneous trajectory of the cantilever provide direct verification of thermal equipartition theorem and the Maxwell-Boltzmann distribution of velocities for the membrane.

The ballistic regime found at short time intervals may be exploited in nanomechanical devices where knowledge of well-defined velocities and of the positions of functional nano-components can give advantage. Moreover, with micro/nano-cantilevers widely used as sensor elements in physical, chemical and biological sciences, short-interval measurements in the ballistic regime present opportunities, such as for fast thermometry, based on evaluation of the initial $k_BT/m_{eff}$ slope of mean squared displacement. Importantly, in being based only on the fundamental rules of thermodynamics and knowledge of the material and geometrical parameters of the cantilever, the calibration of such a thermometry would not require reference to any external standard. The slope measurement may also be used for fast monitoring of cantilever mass at known temperature, for instance during materials deposition processes or for the detection of molecular adsorption and desorption [14]. From a fundamental perspective, the ability to measure the instantaneous velocity of micro/nanomechanical structures will be of importance in the study of non-equilibrium statistical mechanics, for instance in exploring the departure from equipartition theorem expected at the quantum ground state of a cantilever with non-vanishing kinetic energy at 0K. We also argue that the ballistic nature of motion at short time scales should be accounted for in understanding the electrical, thermal and mechanical properties of 2D materials [1-4, 15].

## Methods

Measurements of picometric (sub-atomic scale) cantilever displacement are performed by monitoring of the secondary electron current generated by scattering of a tightly focused electron beam on a sharp edge on the cantilever: This technique relies upon the fact that the secondary electron current $I$ is sensitive to small movements $\delta x(t)$ of the object, giving rise to changes in the current proportional to its gradient in the displacement direction: $\delta I(t) \sim \nabla I \times \delta x(t)$. $I$ and its gradient can be measured in the conventional mode of (static) secondary electron imaging. Fluctuations in cantilever position $\delta x(t)$ can then be measured from variations in current $\delta I(t)$.

In the present case, we used the beam of a scanning electron microscope, with an acceleration voltage of 5 kV and a beam current of 690 pA. For high sensitivity to thermal motion of the cantilever in its fundamental out-of-plane flexural mode (at room temperature under vacuum at 2.6×10⁻⁶ mbar) the electron beam was positioned at the cantilever tip (as indicated in Supplementary Fig. S2a), with the sample plane inclined at 45° to the incident beam direction.

The calibrated time-dependence of cantilever tip displacement (Fig. 1a) is derived from the experimentally recorded secondary electron signal via the measured signal gradient along the movement direction at the sampling location, taking account of the electron beam's angle of incidence relative to the motion direction.

The noise equivalent displacement sensitivity of the technique, which is determined by the sharpness of the membrane edge, size of the focused electron beam spot and shot noise of the secondary electron current, reaches ~1 pm/Hz$^{1/2}$.

**Acknowledgements**

This work was supported by the Engineering and Physical Sciences Research Council, UK (grant numbers EP/M009122/1 and EP/T02643X/1; NIZ, KFM, JYO), the Ministry of Education, Singapore (MOE2016-T3-1-006; NIZ), and the China Scholarship Council (201806160012; TL). The authors would like to acknowledge Neil Sessions' assistance with sample fabrication.

**Data availability**

The data from this paper can be obtained from the University of Southampton ePrints research repository.

# Supplementary Information:
# Ballistic Dynamics of Flexural Thermal Movements in a Nano-membrane Revealed with Subatomic Resolution

Tongjun Liu[1], Jun-Yu Ou[1], Nikitas Papasimakis[1], Kevin F. MacDonald[1], Vitalyi E. Gusev[2], and Nikolay I. Zheludev[1, 3]

[1] Optoelectronics Research Centre and Centre for Photonic Metamaterials, University of Southampton, Highfield, Southampton, SO17 1BJ, UK

[2] Laboratoire d'Acoustique de l'Université du Mans (LAUM), Institut d'Acoustique-Graduate School (IA-GS), CNRS, Le Mans Université, 72085 Le Mans, France

[3] Centre for Disruptive Photonic Technologies, SPMS and TPI, Nanyang Technological University, Singapore, 637378, Singapore

**Micro-cantilever geometry, fabrication, effective mass and mechanical quality factor**

The cantilever employed in this study was 30 nm thick and 62 μm long, with a trapezoidal shape - 3 μm wide at the unclamped end and 0.6 μm wide at the clamped end (Fig. S1). It was manufactured in a freestanding gold film by focused ion beam milling. The gold film was prepared by evaporation of 30 nm gold onto a silicon nitride membrane, with the silicon nitride then removed by reactive ion etching.

For thermomechanical displacement measurements the incident electron beam is positioned at the center of the short edge of the cantilever tip, as indicated in Fig. S2a. Figure S2b shows the power spectral density (PSD) displacement calculated from measured displacement time-series data, clearly revealing thermal motion at the cantilever's fundamental resonance frequency against a noise floor set by secondary electron detection shot noise.

The effective mass $m_{eff}$ and mechanical quality factor $Q$ for the cantilever's fundamental oscillatory mode are obtained by fitting the following analytical expression for displacement power spectral density $S(f)$, from Wiener–Khinchin theorem [1, 2] to the experimental data:

$$S(f) = \frac{k_B T f_0}{2\pi^3 m_{\text{eff}} Q[(f_0^2 - f^2)^2 + (ff_0/Q)^2]}$$

This fitting – the orange curve in Fig. S2b – yields values $m_{eff}$ = 47 pg (~2/3 the cantilever's estimated 72 pg real mass, from dimensions and the density of gold), and $Q$ = 501.

**Disturbance of cantilever movement by the probe electron-beam**

The increase in cantilever temperature induced by injection of probe electrons has been evaluated via numerical modelling. Monte Carlo simulations [3] show that >95% of 5keV electrons are stopped within the 30 nm thickness of the beam. In keeping with prior works [4] we assume that around 2.5% of incident electron beam power is absorbed as heat $H$ = 0.025$IV$ ~100 nW. Taking the specific heat capacity of gold to be 700 Jkg$^{-1}$K$^{-1}$, the cantilever temperature is increased by electron bombardment by $\delta T$ ~ 0.4 K resulting in a negligible relative increase of the root mean square thermal displacement of the cantilever by only one part in ($\sqrt{1 + \delta T/T} - 1$) ~7×10$^{-4}$.

The force related to momentum transfer from the electron beam is $F = \frac{I}{e}\sqrt{2Em_e}$ ~1.6×10$^{-13}$ N. Assuming a spring constant for the cantilever of 60 μN/m (from finite element modelling), this is sufficient to induce static bending (i.e. tip displacement) of ~2.6 nm. This is much smaller than the (30 nm) thickness of the cantilever and of no consequence to its thermal motion dynamics.

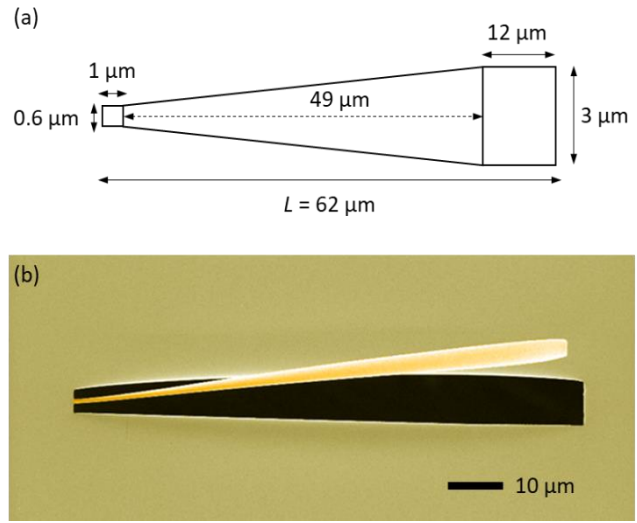

**Fig. S1. Nanomechanical micro-cantilever.** (a) In-plane dimensional schematic of the cantilever. (b) False color scanning electron microscope image of the cantilever, taken at a 30° viewing to the free-standing gold membrane surface normal.

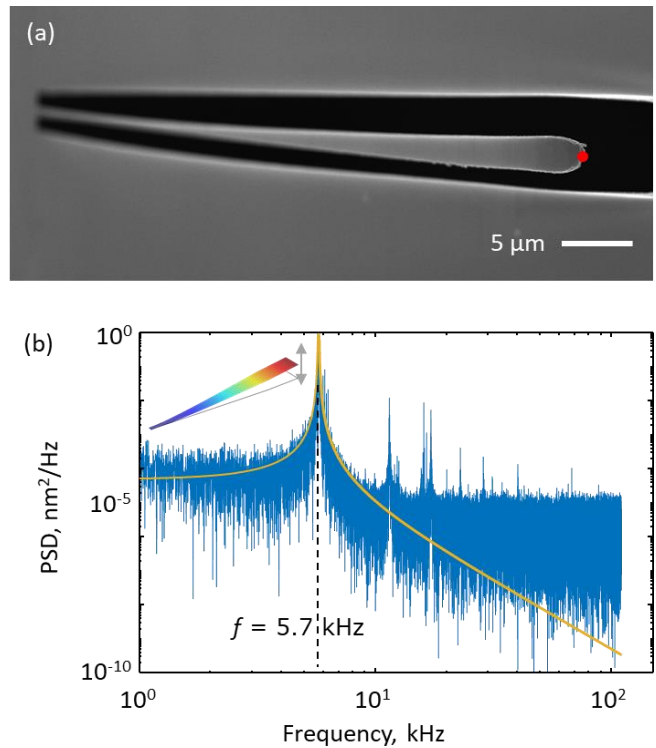

**Fig. S2. Displacement power spectral density at the cantilever tip.** (a) Scanning electron microscope image of the cantilever taken in the orientation in which tip-displacement time-series measurements are performed, i.e. with the sample plane inclined at 45° to the incident electron beam [in consequence of which, the left hand side of the image is out of focus]. (b) Displacement power spectral density (PSD) of the cantilever tip measured at the position denoted by the red dot in panel (a). The overlaid orange line is a best fit of the analytical expression for PSD given in Eq. S1.